\documentclass[10pt]{iopart}

\usepackage{mathrsfs}
\usepackage{iopams}
\usepackage{cite}
\usepackage{epsfig}
\usepackage{graphicx}
\usepackage{color}

\def\eps{\epsilon}
\def\Th{\Theta}

\def\3nab{\tilde{\nabla}}
\def\R{\tilde{R}}

\def\hsp5{\hspace{5mm}}
\newcommand{\sfrac}[2]{{\textstyle{#1\over#2}}}
\def\case#1/#2{\textstyle\frac{#1}{#2}}

\def\be {\begin{equation}}
\def\ee {\end{equation}}
\def\bea {\begin{eqnarray}}
\def\eea {\end{eqnarray}}

\def\case#1/#2{\textstyle\frac{#1}{#2} }
\def\rf#1{(\ref{#1})}

\def\equi {equilibrium\ }
\def\ES {Einstein Static\ }

\def\nt {\tilde{n}}
\def\D {\tilde{\nabla}}

\def\cqg{{\em Class. Quantum Grav.\/} }
\def\grg{{\em Gen. Rel. Grav.\/} }
\def\prd{{\em Phys. Rev.\/} {\bf D}}
\def\prl{{\em Phys. Rev. Lett.\/} }
\def\apj{{\em Astrophys. J.\/} }
\def\jmp{{\em J. Math. Phys.\/} }
\def\mn{{\em Mon. Not. Roy. Astr. Soc.\/} }
\def\aph{{\em Ann. Phys. (NY)\/} }
\def\plb{{\em Phys. Lett.\/} {\bf B}}

\begin{document}
\title{Dynamics of $f(R)$-cosmologies containing Einstein static models}
\author{Naureen Goheer \dag, Rituparno Goswami \dag \ and Peter K.S. Dunsby \dag \ddag}
\address{\dag \ Department of Mathematics and Applied Mathematics, University of Cape Town, Rondebosch,
7701, South Africa}

\address{\ddag \  South African Astronomical Observatory, Observatory, Cape Town, South Africa}

\date{\today}

\eads{naureen.goheer@gmail.com, Rituparno.Goswami@uct.ac.za and 
peter.dunsby@uct.ac.za}

\begin{abstract}
We study the dynamics of homogeneous isotropic FRW cosmologies with positive 
spatial curvature in $f(R)$-gravity, paying special attention to the existence of 
Einstein static models and only study forms of $f(R)=R^n$ for which these static 
models have been shown to exist. We construct a compact state space and identify past and
future attractors of the system and recover a previously discovered future attractor 
corresponding to an expanding accelerating model. We also discuss the existence of universes
which have both a past and future bounce, a  phenomenon which is absent in General Relativity.  
\end{abstract}

\section{Introduction}
After more than one hundred years, Einstein's theory of General Relativity still remains
our best description of gravity and has up to now survived the scrutiny of a multitude 
of tests, most of which have been on Solar System scales. However, in the last decade, high 
precision astrophysical and cosmological observations appear to suggest that  General Relativity 
might be incomplete. In particular, cosmological data indicates an underlying cosmic acceleration of the
Universe cannot be recast in the framework of General Relativity without resorting to additional exotic 
matter components (known as Dark Energy), which have not yet been directly observed. Several models have
been proposed  \cite{lambda-darksector} in order to address this problem and at the moment,  the one 
which appears to fit all available observations (Supernovae Ia \cite{sneIa}, Cosmic Microwave
Background Anisotropies \cite{cmbr}, Large Scale Structure formation \cite{lss}, baryon oscillations 
\cite{baryon}, weak lensing \cite{wl}), turns out to be the so called {\it Concordance Model} in
which a tiny cosmological constant is present \cite{astier} and ordinary matter is dominated by 
Cold Dark Matter (CDM). However, despite its success, the $\Lambda$\,-\,CDM model is affected by 
significant fine-tuning problems related to the vacuum energy scale, so  it is important to investigate other 
viable theoretical schemes.

Currently, one of the most popular alternatives to the {\it Concordance Model} is based on modifications of 
standard Einstein gravity.  Although modifications of Einstein's theory of  gravity were already proposed in the 
early years after the publication of General Relativity, a detailed investigation of cosmological models within 
this framework only got underway a few years ago. Such models became popular in the 1980's because it was shown 
that they naturally admit a phase of accelerated expansion which could be associated with an early universe inflationary 
phase \cite{star80}. The fact that the phenomenology of Dark Energy requires the presence of a similar phase (although 
only a late time one) has recently revived interest in these models. In particular, the idea that Dark Energy may have a 
geometrical origin, i.e., a that there is a connection between Dark Energy  and a non-standard behavior of gravitation on 
cosmological scales is currently a very hot topic of research.

Among these models the so-called  Extended Theory of Gravitation (ETG) and, in particular, 
{\em higher-order theories of gravity} (HOG)  \cite{DEfR} have provided a number of extremely interesting
results on both cosmological  \cite{ccct-ijmpd,review,cct-jcap,otha} and astrophysical 
\cite{cct-jcap,cct-mnras} scales. These models are based on gravitational actions which are 
non-linear in the Ricci curvature $R$ and$/$or contain terms involving combinations of derivatives of 
$R$ \cite{kerner,teyssandier,magnanoff}. One of the nice features of  these theories is that 
the field equations can be recast in a way that the higher order  corrections are written as an 
energy\,-\,momentum tensor of geometrical origin describing an ``effective" source term on the right hand 
side of the standard Einstein field equations  \cite{ccct-ijmpd,review}. In this {\em Curvature Quintessence} scenario, 
the cosmic acceleration can be shown to result from such a new geometrical contribution to the cosmic energy density 
budget,  due to higher order corrections of the Hilbert-Einstein Lagrangian. 

Unfortunately the analysis of HOG is complicated by the fact that the resulting fourth-order field equations are 
highly non-linear and finding cosmological solutions by solving the field equations directly has proved to be 
extremely difficult. This problem has been eased somewhat by using the theory of dynamical systems which 
has over the last few years proved to be a powerful scheme for investigating the physical behavior of such 
theories (see for example \cite{cdct:dynsys05,ScTnDynSys,Carloni1,Leach1,Leach2,Goheer1,Goheer2,Shosho}). 
In fact, studying cosmologies using the dynamical systems approach has the advantage of providing a relatively 
simple method for obtaining exact solutions (even if these only represent the asymptotic behavior) and  obtain a 
(qualitative) description of the global dynamics of these models. Consequently, such an analysis allows for an 
efficient preliminary investigation of these theories, suggesting which background models deserve further investigation
and this provides a starting point for the analysis of the growth of structure in HOG \cite{Carloni2, Ananda}.

A remarkable feature of the background dynamics of  $f(R)$ gravity is that there are a number of classes of these 
theories that admit a Friedmann  transient  matter-dominated decelerated expansion phase, followed by one 
with an accelerated expansion rate \cite{cdct:dynsys05,Capozziello:2006dj}. The first phase provides a setting 
during which structure formation can take place and this is followed by a smooth transition to a Dark Energy like 
era which drives the cosmological acceleration. It would therefore be of great interest if orbits could be found in the 
phase-space of $f(R)$ models that connect such a Friedmann matter dominated phase to an accelerating phase 
via an Einstein Static solution in a way which is indistinguishable (at least at the level of the background dynamics) 
to what occurs in the $Lambda$CDM model of General Relativity \cite{Goliath}.  If a similar evolution exists in an $f(R)$ model,
we would expect the associated Einstein Static solution to be a saddle point (as it is in General Relativity) and consequently unstable. 
It therefore follows  that a careful examination of the existence and stability of the Einstein Static model in the more general setting of modified gravity 
is of  critical importance in determining whether $\Lambda$CDM - like cosmologies are a generic feature of such gravitational theories. 

In General Relativity, the issue of  stability of the Einstein Static model has been studied several times since the classic paper by
Eddington \cite{eddington} in the 1930s, where it was shown that such models are unstable with respect to homogeneous 
and isotropic perturbations, exactly the feature that allows a transition between a decelerated expansion era to one which 
is accelerating. However, later work by Harrison \cite{harrison} and Gibbons \cite{gibbons}, which extended Eddington's work, 
considered generic inhomogeneous and anisotropic perturbations of an Einstein Static model filled with a perfect fluid 
and found that  provided the sound speed satisfies $c_{{\rm s}}^{2} >{\frac{1}{5}}$, these models are marginally stable with 
respect to such perturbations. The reason for this ``non-Newtonian"  stability stems from the fact that in General Relativity, the Einstein Static
universe is spatially closed, and therefore has a maximum scale associated with it, which is greater than the largest physical scale of the 
perturbations. Since the Jean's length is a significant fraction of this maximum scale, perturbations in the fluid oscillate, rather than 
grow, leading to the conclusion that in General Relativity at least,  the Einstein Static solution is marginally stable with respect to such perturbations. 
This result was also found more recently by Barrow {\it et. al.} \cite{barrow03}

Recently  \cite{Goswami:2008fs} the existence and stability of the Einstein Static models in the more general setting of $f(R)$ gravity was examined.
It was found that only one class of $f(R)$ theories admits an Einstein Static model, and that this class is neutrally stable with respect to vector 
and tensor perturbations for all equations of state on all scales. Scalar  perturbations are only stable on all scales if the matter fluid 
equation of state satisfies $c_{{\rm s}}^{2}>\frac{\sqrt{5}-1}{6}\approx 0.21$. This result is remarkably similar to the GR  case discussed above. 

In this paper we use the theory of dynamical systems to examine in detail the background evolution of this special class of $f(R)$ theories. 
This analysis compliments the work in  \cite{Goswami:2008fs} and provides a more complete picture of how the Einstein Static solution fits into the 
overall structure of the phase space of solutions in $f(R)$ gravity.

The following conventions will be used in this paper: the metric signature is $(-+++)$; Latin indices run from 0 to 3; units are used
in which $c=8\pi G=1$.
\section{The Field Equations}
For homogeneous and isotropic spacetimes, a general form of the
action for fourth-order gravity is given by
\begin{equation}
{\cal A}=\int dx^4 \sqrt{- g}\left [f(R)-2\Lambda +{\cal L}_m\right],
\label{action:f(R)}
\end{equation}
where ${\cal L}_m$ is the Lagrangian of the matter fields. The
fourth-order field equations can be obtained by varying
\rf{action:f(R)}: 
\begin{equation}
\label{field:f(R)T}
G_{ab}+g_{ab}\frac{\Lambda}{f'}=\frac{T_{ab}^m}{f'}+ T_{ab}^R\equiv
T_{ab}^T\;,
\end{equation}
where primes denote derivatives with respect to $R$. Here $T_{ab}^T$
is the total effective energy momentum tensor composed of the
ordinary matter energy momentum tensor $T_{ab}^m$ and the correction
term $T_{ab}^R$ (often referred to as the "curvature fluid"):
\begin{equation}
T_{ab}^R=\frac{1}{f'}\left[\frac{1}{2}g_{ab}(f-Rf')+
f'_{;cd}(g^c_ag^d_b-g^{cd}g_{ab})\right]\,.
\end{equation}
As shown in \cite{Goswami:2008fs}, if we wish to keep the extra degrees of freedom 
in fourth-order gravity, the existence of an Einstein static universe
imposes a strong constraint on the function $f(R)$, which was found to be
\begin{equation}
f(R)=2\Lambda+{\cal K}R^{\frac{3}{2}(1+w)}\,\, , \,w\neq -1.
\end{equation}
Here we assume that the standard matter is a barotropic perfect fluid 
with equation of state 
\begin{equation}
p^m=w\rho^m\,,
\end{equation}
where $\rho^m$ and $p^m$ are the standard matter density and pressure respectively.
The effective field equations will look like the ones for $R^n$--gravity, 
but subject to the constraint
\begin{equation}
\label{con-n-w} n=\frac{3}{2}(1+w)\,,
\end{equation}
which guarantees the existence of an Einstein Static solution.
This means the equation of state is fixed as a function of $n$, or
in other words: for a given value of $n$ we can only find Einstein
static solutions if the equation of state satisfies (\ref{con-n-w}).
As we know, in order to satisfy all the energy conditions for standard matter, 
$w$ has to take values in $[-1/3,1]$. We will therefore only
consider the range $n\in [1,3]$, and in particular will analyze in detail 
the cases of dust ($n=3/2,~w=0$) and radiation ($n=2,~w=1/3$).
In what follows we will use equation (\ref{con-n-w}) to
eliminate $w$ from the field equations.

For the homogeneous and isotropic spacetimes, the independent field equations 
for $R^n$--gravity are (with $n$ given by (6):
\begin{itemize}
\item
The Raychaudhuri equation 
\begin{equation}
\dot{\Th} + \sfrac{1}{3}\,\Th^{2}  -\frac{1}{2n}R -
(n-1)\frac{\dot{R}}{R}\Th+\frac{\rho^m}{nR^{n-1}} =0,
\label{Ray:R^n_BOrt}
\end{equation}
where $\Theta$ is the volume expansion which defines a scale factor
$a(t)$ along the fluid flow lines via the standard relation
$\Theta=3\dot{a}/{a}$. 
\item
The Friedmann equation 
\begin{equation}\label{fried}
\Th^{2}+\frac{3}{2}\R=-3(n-1)\frac{\dot{R}}{R}\Th+
3\frac{(n-1)}{2n}R+\frac{3\rho^m}{nR^{n-1}}\,,
\end{equation}
where $\R\equiv (6\kappa)/a^2$ is the curvature of the 3-spaces and 
$\kappa\in(1,0,-1)$ denotes closed, flat and open universes respectively.  
\item
The trace equation 
\begin{equation}
\frac{\ddot{R}}{R}=\frac{1}{3}\frac{(n-2)}{n(n-1)}R -\theta\dot{R}-(n-2)\frac{\dot{R}^2}{R^2}
+\frac{2}{3}\frac{(2-n)}{n(n-1)}\frac{\rho^m}{R^{n-1}}\;.
\label{trace}
\end{equation}
\item
The conservation equation for standard matter and the propagation equation for 
the 3-curvature are
\begin{equation}\label{cons:perfect}
\dot{\rho}^m=-\frac{2}{3}n\rho^m\Th\;\;,\;\;\dot{\R}=-\frac{2}{3}\R\Th
\end{equation}
respectively. 
\end{itemize}
Combining the Raychaudhuri and Friedman equations, we get
\begin{equation}\label{thetadot}
R=2\dot{\Th}+\frac{4}{3}\Th^2+\R\;,
\end{equation}
which is equivalent to the general definition of the Ricci scalar in terms of 
the scale factor in a FRW spacetimes, given by
\begin{equation}
R=6\left(\frac{\ddot{a}}{a}+\frac{\dot{a}^2}{a^2}+\frac{\kappa}{a^2}\right)\;.
\label{R}
\end{equation}
For our analysis of the dynamical system, which will be presented in the next section, it is useful to complete the square 
of equation (8) and rewrite the Friedmann equation as
\begin{equation}
\left(\Th + \frac{3(n-1)}{2}\frac{\dot{R}}{R}\right)^2+\frac{3}{2}\R=
\frac{9(n-1)^2}{4}\frac{\dot{R}^2}{R^2}+3\frac{(n-1)}{2n}R+\frac{3\rho^m}{nR^{n-1}}\,.
\label{fried2}
\end{equation}
We can easily see that all the additive terms on both sides of the equation above 
are strictly non-negative for all spacetimes with non-negative Ricci scalar $R$ and 3-curvature $\R$. 
This will help us to compactify  the full FRW state space for non-negative spatial curvature and Ricci scalar, 
and thus allow us to analyze the dynamics of the system in a compact framework.
\section{Dynamics of non-negative curvature FRW spacetime}
We now study the dynamics of FRW models with non-negative spatial curvature 
and Ricci scalar, which include the Einstein static universe. 
In order to convert the equations above into a system of autonomous first order
differential equations, we define the following set of 
normalized variables \footnote{It is important to note that this
choice of variables will exclude General Relativity, i.e., the case of $n=1$. See
\cite{Goliath} for the dynamical systems analysis of the corresponding cosmologies in GR.};
\begin{eqnarray}\label{DS_orth:var}
&&x = \frac{3\dot{R}}{2R D}(n-1)\; ,\quad y = \frac{3R}{2n
D^2}(n-1)\; ,\quad  z = \frac{3\rho^m}{nR^{n-1}D^2}\; ,
\\    
&&K=\frac{3\R}{2D^2}\;  ,\quad\quad Q=\frac{\Theta}{D} , \nonumber
\end{eqnarray}
together with the normalized time variable
\begin{equation}\label{def-dash}
'\equiv \frac{d}{d\tau}\equiv \frac{1}{D}\frac{d}{dt}\,.
\end{equation}
Here the normalization $D$ is chosen as
\begin{equation}\label{D1}
D\equiv \sqrt{\left(\Theta+ \frac{3(n-1)}{2}\frac{\dot{R}}{R}\right)^2+\frac{3}{2}\R}
\end{equation}
in order to compactify the variables as shown below.
In terms of these variables, the Friedmann equation (\ref{fried2}) becomes
\begin{equation}
x^2+y+z=1\;,~ y,z\geq0\;,
\end{equation}
and from the definition of the normalization we get another constraint
\begin{equation}
(Q+x)^2+K=1\;,~K\geq0.
\end{equation}
The above constraints show that the variables must be compact and take the following ranges:
\begin{equation}
x\in[-1,1]\;,y\in[0,1]\;z\in[0,1]\;,Q\in[-2,2]\;,K\in[0,1]\;.
\end{equation} 
Using the five compact variables together with the two constraints, we reduce the complete 
dynamical system to a 3-dimensional one, and the cosmological equations become equivalent 
to the autonomous system
\begin{eqnarray}\label{DS1}
Q'&=\left[(3-n)x^2-n(y-1)-1\right]\frac{Q^2}{3}+\left[(3-n)x^2-n(y-1)+1\right]\frac{Qx}{3}
\nonumber\\
&+\frac{1}{3}\left[x^2-1+\frac{ny}{n-1}\right]\;, \nonumber\\
\label{DS}
y'&=\frac{2yx^2}{3}(3-n)(x+Q)+\frac{2xy}{3}\left[\frac{(n^2-2n+2)}{n-1}-ny\right] 
\nonumber\\
&+\frac{2}{3}Qny(1-y)\;,\\
x'&=\frac{x^3}{3}(3-n)(Q+x)+\frac{x^2}{3}\left[n(2-y)-5\right]\nonumber\\
&+\frac{Qx}{3}\left[n(1-y)-3\right]+\frac{1}{3}\left[\frac{n(n-2)}{n-1}-n+2\right]\;.\nonumber
\end{eqnarray}
\section{Qualitative Analysis of the state space}
The complete description of all the equilibrium points of the dynamical system described by the 
above equations is given in Table I. We note that the points $\mathcal{A}$ and $\mathcal{B}$ correspond 
to the similarly labeled points in \cite{goheer}. The line $\mathcal{LC}$ has been 
labeled as such since it  includes the point $\mathcal{C}$ found in \cite{goheer} 
in the flat limit  $|Q+x|=|(2-n)\,Q|\rightarrow 1$.  Strictly speaking the system (\ref{DS})
has an additional equilibrium point, but its coordinates only lie in the state space 
for $n>3$, which for us corresponds to an unrealistic equation of state $w>1$. 
Note that the subscript $\eps$ only labels expanding ($\eps=+1$) and collapsing ($\eps=-1$) 
models for the point $\mathcal{A}$. Since the other points correspond to 
asymptotically Minkowski spaces, the subscript merely labels the different 
limits from which these asymptotic solutions were obtained. The line $\mathcal{LC}$ 
has an expanding branch  $\mathcal{LC}^{\rm exp}$ with $Q>0$, a collapsing branch 
$\mathcal{LC}^{\rm coll}$ with $Q<0$, and in between the Einstein Static model  
$\mathcal{ES}$, with $Q=0$. The stability  of the equilibrium points and the line is 
summarized in Table II. We will now briefly discuss several interesting features 
of these equilibrium points.

\begin{enumerate}  
\item The position of the \equi points $\mathcal{N}_\eps$ in the state space 
does not depend on the equation of state for the standard matter. 
In fact,  the existence of these points is a generic feature 
of $R^n$ gravity, and they form a global source ($\eps=+1$) or sink ($\eps=-1$) in the state space considered here. 
This implies that for any equation of state, there always exists the possibility 
of the Universe originating from a vacuum Minkowski model in the infinite past, or 
evolving towards a vacuum Minkowski model in the infinite future.

\item The position of the \equi points $\mathcal{L}_\eps$ also does not depend 
on the equation of state for standard matter. However, the stability properties of these 
points do depend on the equation of state. For matter with high negative pressure $(-1/3\le w\le -1/6)$, 
or for very stiff matter $(2/3\le w\le 1)$, these points are unstable saddles. 
For other values of the barotropic index, these points represent a  global source/sink depending 
on the sign of $\eps$.

\item The points $\mathcal{B}_\eps$ only exist for the specific range of the 
barotropic index $w \in [-1/3, 2/3]$, and they are always unstable.

\item The points $\mathcal{A}_\eps$ exist for $w\in [-1/6, 1]$ only, and they have the interesting 
property of exhibiting de Sitter (dS)/anti-de Sitter (AdS) like acceleration/deceleration depending on 
the barotropic index. In this case however, the acceleration has a power law behavior instead of 
the exponential behavior of dS/AdS. These points are vacuum solutions,  and the expanding point is a global sink 
if $n>P_+\approx 1.37$ or $w>w_+\approx -0.09$. These points 
are also a generic feature of $R^n$ gravity, and  
the existence of these points implies that even with an ultra-violet correction to General Relativity, one 
can evolve towards late time acceleration for the ranges of $w$ given in Table I. 

\item The line $\mathcal{LC}$ is entirely an artifact of the $n-w$ correspondence 
(\ref{con-n-w}), and is absent in a general $R^n$-gravity. All the points on this 
line are expanding/collapsing non-accelerating solutions, containing the Einstein static 
solution as the only static point. For $n\in [1,P_+]\approx [1,1.37]$ 
(see table caption), the expanding branch of  the line forms spiral sinks, 
while the collapsing branch of the line forms spiral 
sources. The Einstein static point in between these branches forms a center, that 
means it is neutrally stable for this range of $n$. For $n\in [P_+, 3/2]$ 
a bifurcation appears on the line: for $Q\in[-Q_b,Q_b]$, the expanding (contracting) 
points on the line remain spiral sinks (sources), and the Einstein static point 
remains a center. The points with $|Q|>Q_b$ however now are saddles.  
Here $Q_b$ is the bifurcation value defined by 
\begin{equation}
Q_b\equiv \sqrt{\frac{2n-3}{(n-1)(11-11n+2n^2)}}\,.
\end{equation}
We can see that the bifurcation enters the state space from both endpoints of the 
lines for $n=P_+$. As $n$ increases, $Q_b$ and $-Q_b$ move closer towards the center 
of the line, and for $n=3/2$ the two bifurcations $\pm Q_b$ merge at $Q=0$.   
For equations of state stiffer than dust ($n>3/2$), all the points on the 
line are saddles, including the Einstein static model.
  
\item The state space in general contains several past and future attractors. 
Interestingly, we find that for any equation of state, there is no expanding 
past attractor. The only possible past attractors are the asymptotically Minkowski 
points  $\mathcal{L_+}$ and  $\mathcal{N_+}$, and the decelerating collapsing point  
$\mathcal{A}_-$. For $w<0$, the  points on the collapsing non-decelerating  
line $\mathcal{LC}^{\rm coll}$ may also be past attractors.   
This  means that in this class of models, there is 
\textit{no Big Bang scenario} for any \textit{realistic equation of state} 
$w\in [-1/3,1]$, but only possible bounce scenarios or expansion after an 
initial asymptotic Minkowski phase. Figure I shows the state space for dust ($n=3/2$), 
with all the equilibrium points. There are bounce scenarios along those 
trajectories where the quantity $Q$ changes sign. Figure 4 shows a solution which is bouncing 
in past as well as in the future. These kind of solutions are not present in 
General Relativity.  
\end{enumerate}

\begin{figure}[tb]
\begin{center}
\includegraphics[width=11cm,height=9cm]{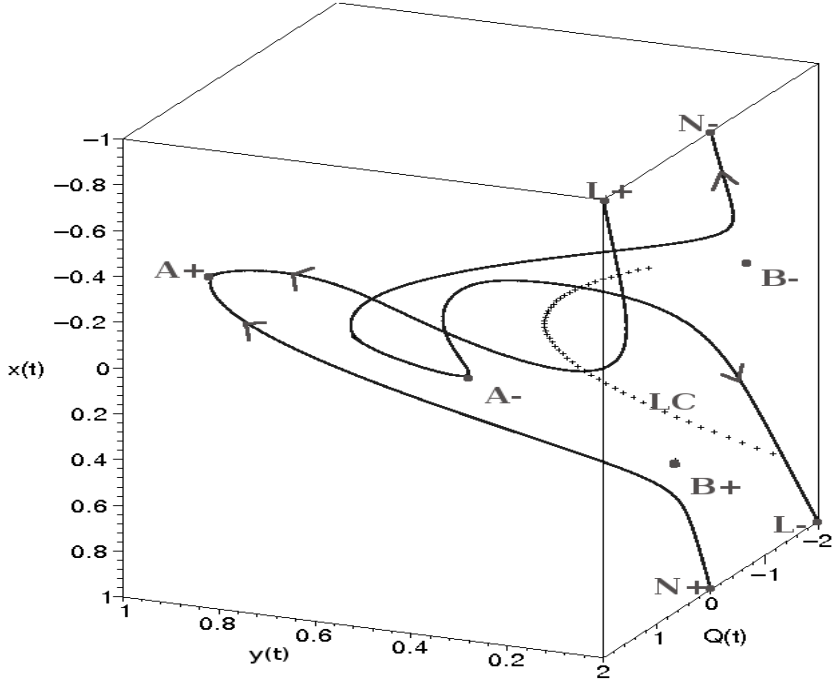}
\caption{The complete state space for dust with all the equilibrium points. We can see the 
trajectories coming out from the global sources $\mathcal{L}_+$, $\mathcal{N}_+$ and 
$\mathcal{A}_-$ and going to the sinks $\mathcal{L}_-$, $\mathcal{N}_-$ and 
$\mathcal{A}_+$. Along the trajectory from $\mathcal{A}_-$ to $\mathcal{L}_-$ the quantity 
$Q$ changes sign twice which shows bounce scenarios in past and future.}
\end{center}
\end{figure}

\begin{figure}[tb]
\begin{center}
\includegraphics[width=11cm,height=9cm]{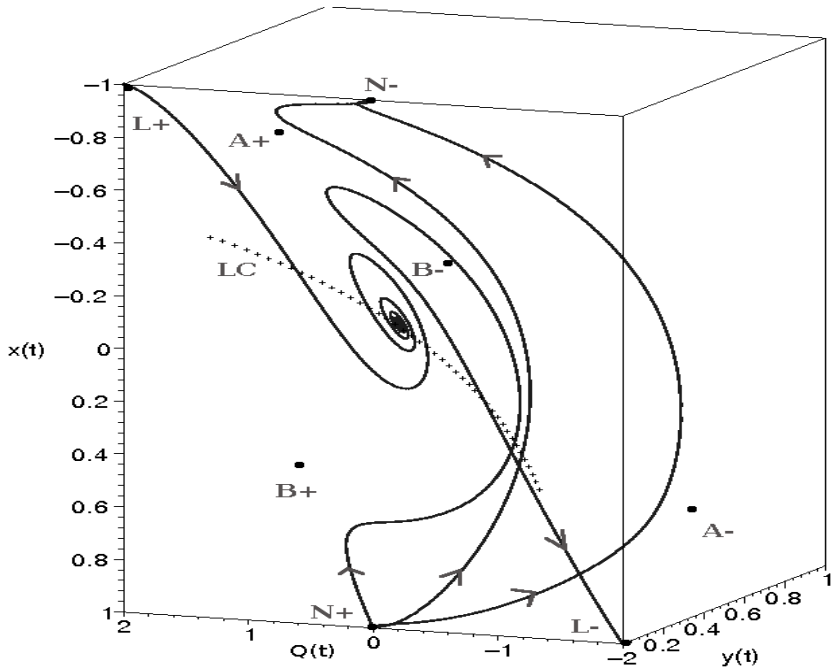}
\caption{The complete state space for matter with $w=-1/9$,  with all the equilibrium points. 
The expanding part of line ${\mathcal{LC}}$ now behaves as a spiral sink. Also the points
$\mathcal{A}_-$ and $\mathcal{A}_+$ are now saddles.}
\end{center}
\end{figure}

\begin{figure}[tb]
\begin{center}
\includegraphics[width=11cm,height=9cm]{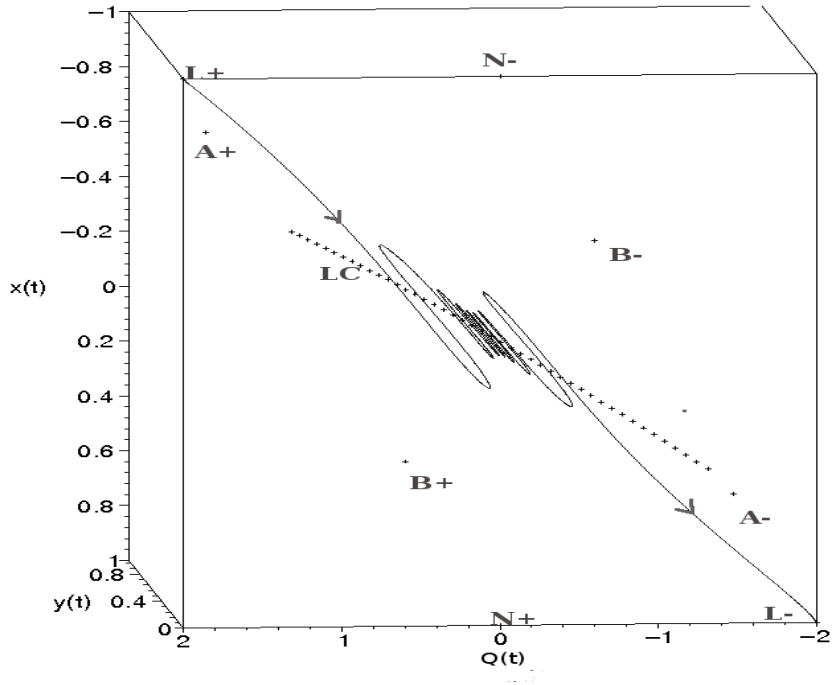}
\caption{The quasiperiodic nature of the \ES point for $w=-1/9$. Here the initial conditions 
are taken very close to the \ES point on the plane perpendicular to the line 
$\mathcal{LC}$ containing the \ES point.}
\end{center}
\end{figure}

\begin{figure}[tb]
\begin{center}
\includegraphics[width=5cm,height=6cm]{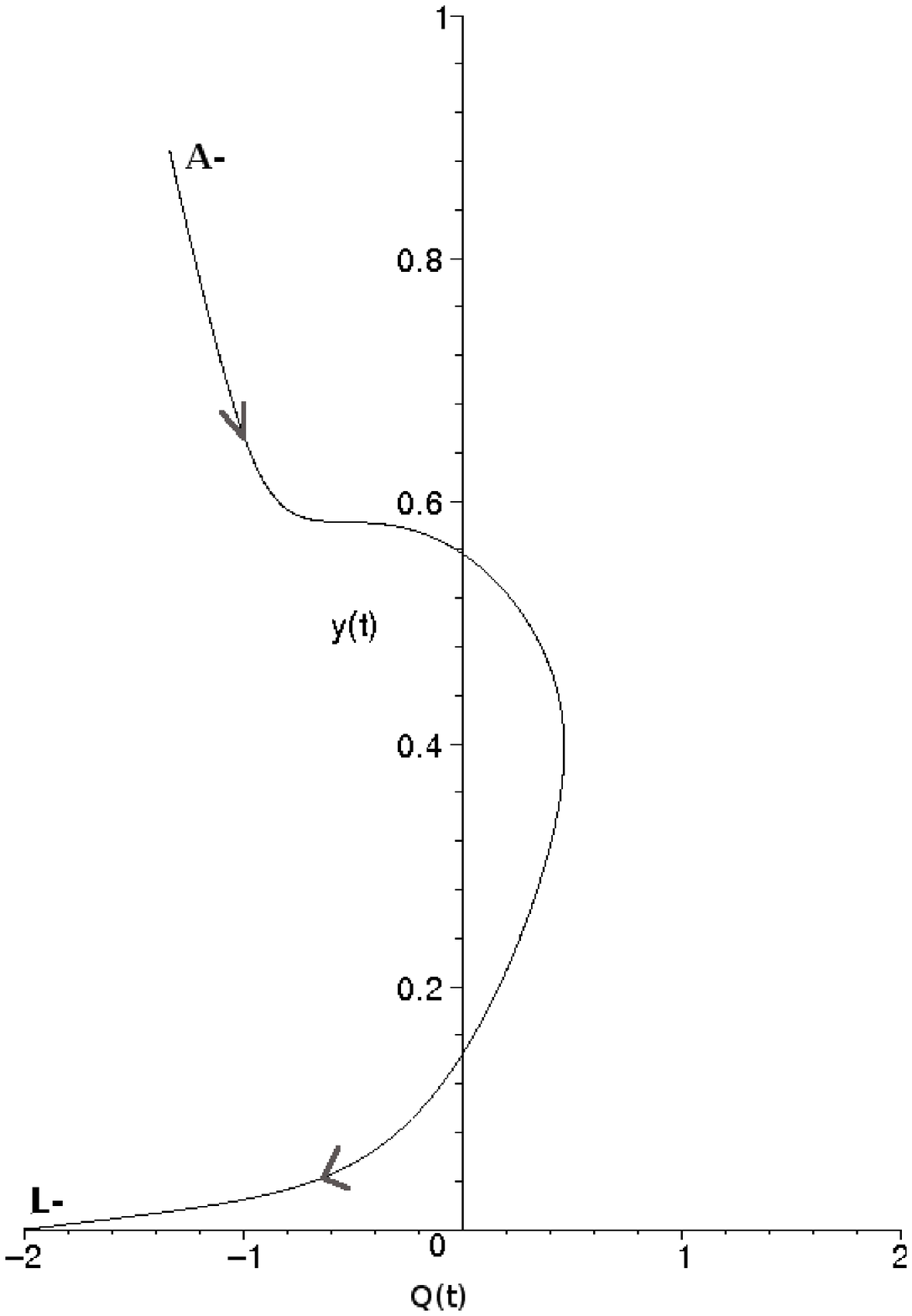}
\includegraphics[width=5cm,height=6cm]{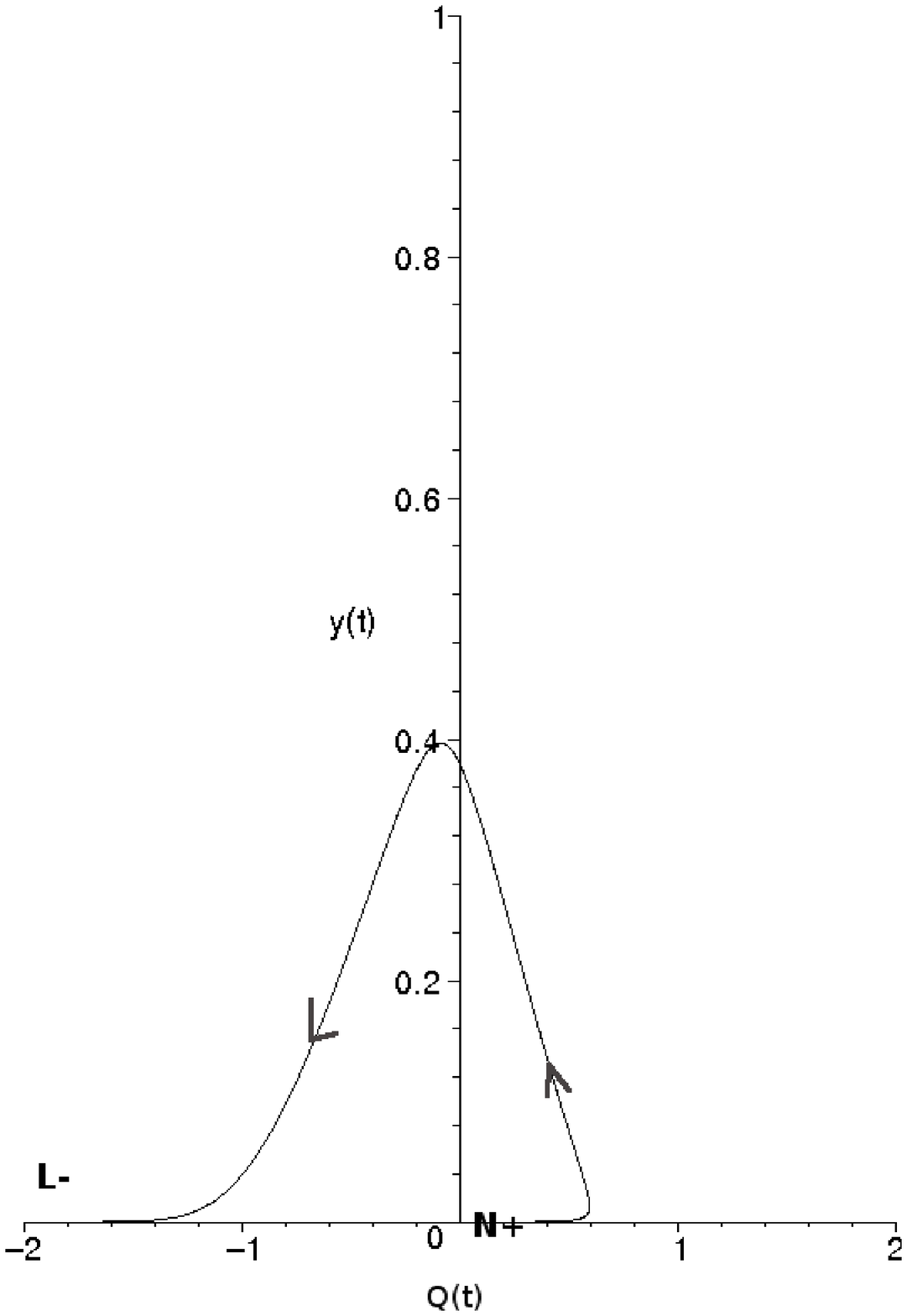}
\caption{Example of bouncing solutions for dust (left) and radiation (right). The dust solution
shows a bounce in both past and future which is not possible in General Relativity.}
\end{center}
\end{figure}

\begin{figure}[tb]
\begin{center}
\includegraphics[width=8cm,height=8cm]{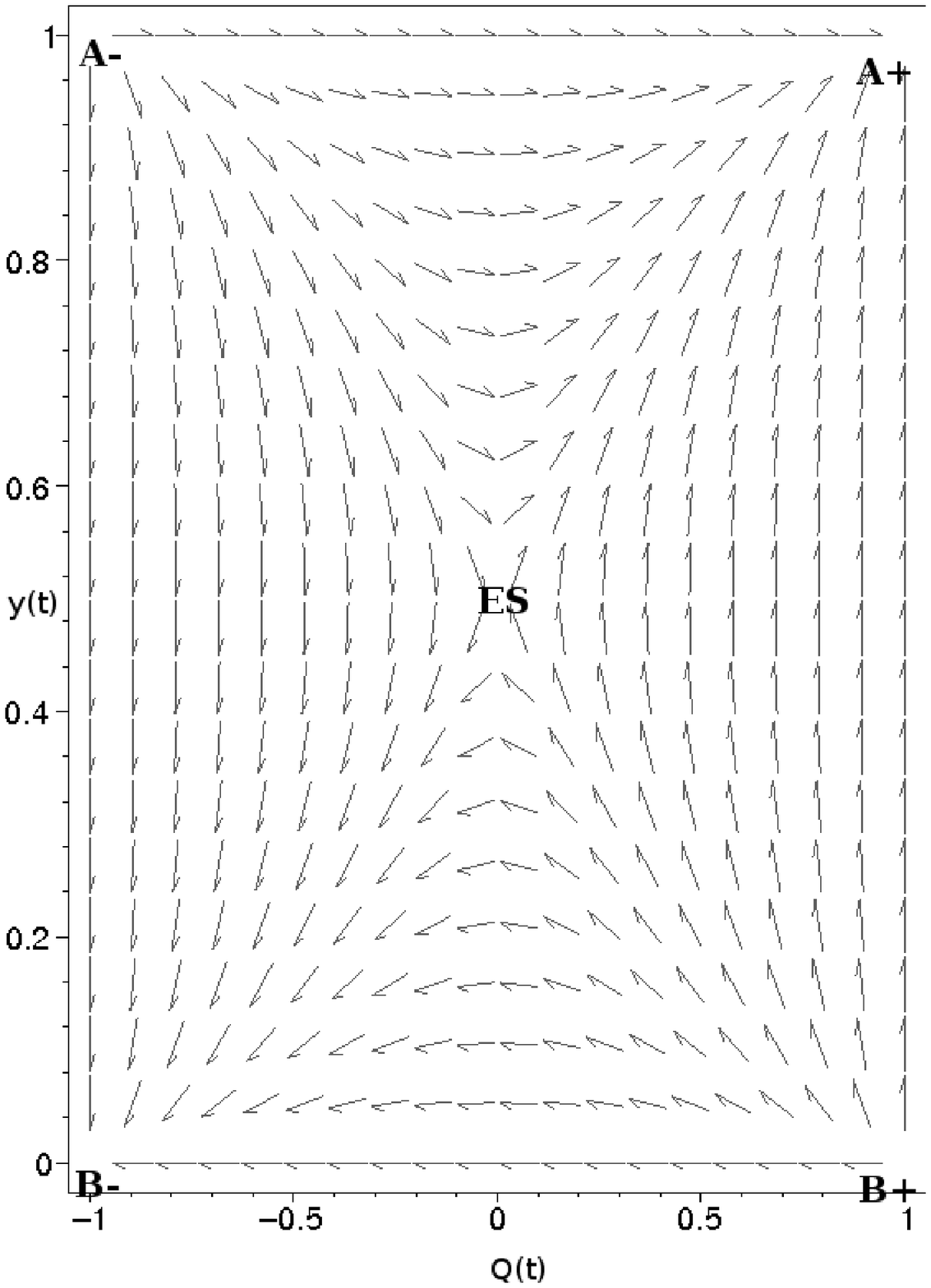}
\caption{The $x=0$ invariant subspace for radiation, containing the points 
$\mathcal{A}_\eps$, $\mathcal{B}_\eps$  and $\mathcal{ES}$. It is interesting to note that in 
this subspace the points $\mathcal{B}_\eps$ behaves like source/sink, however in the complete 
state space they are saddles.}
\end{center}
\end{figure}

\subsection{Stability of the Einstein Static model}

We will now discuss the stability of the Einstein static point in more detail. 
The stability of the \ES model against general covariant, gauge-invariant 
linear perturbations was studied in \cite{Goswami:2008fs}. 
The standard procedure of harmonic decomposition is used, employing 
the trace-free symmetric tensor eigenfunctions $Q$ of the spatial Laplace-Beltrami operator:
\be
\D^2Q=-\frac{k^2}{a_0^2}Q\;,\; \dot{Q}=0\;,
\ee
where the constant $a_0$ defines the length scale for \ES universe. 
For the spatially closed models, the spectrum of eigenvalues is
discrete and given by \cite{harrison,BDE}
\be
k^2=\nt(\nt+2)\;,
\ee
where the co-moving wavenumber takes the values $\nt=1,2,3,\cdots$.
We note that the dynamical systems analysis discussed in the present paper 
corresponds to the homogeneous mode with comoving wavenumber $\nt=0$ 
in the harmonically decomposed perturbation equations. 
This mode was not studied in \cite{Goswami:2008fs}, since it corresponds to a 
change in the background of the linear perturbation theory.

It can be easily seen from equation (53) in \cite{Goswami:2008fs} that the scalar 
perturbations for $\nt=0$ have a growing and a decaying mode if $w>0$. 
If $w<0$ on the other hand, the density gradient has two purely oscillatory modes. 
These results agree with the stability of the \ES model in the dynamical systems 
context analyzed here, where we found that the \ES model is a saddle for $w>0$. 
For $w<0$, since all the points of the line $\mathcal{LC}$ at one side of \ES are sources
and other side of \ES are sinks, the point \ES becomes quasiperiodic in nature, in the sense 
that orbits emerge from a source, spiral around \ES for some time and goes to a sink.
Figure 3 shows such an orbit, where the initial conditions are taken very close 
to the \ES point on the plane perpendicular to the line $\mathcal{LC}$ containing 
the \ES point. This quasiperiodic nature reflects in the oscillatory modes of 
the scalar perturbations in \cite{Goswami:2008fs}.  

As we can easily see from the dynamical system equations, for dust-like matter 
$(w=0)$ the $\mathcal{ES}$ point is a double bifurcation point, and consequently 
the linear theory breaks down. The physical interpretation is as follows: from equation 
(53) in 
\cite{Goswami:2008fs} we can easily see that for $w=0$ and $\nt=0$, the linear perturbation 
is independent of time, i.e., it is a constant. Hence to study the actual behavior 
of the homogeneous perturbation for dust, we must consider higher order corrections.

\subsection{The state space for radiation}

For $n=2$ (or $w=1/3$) the state space has some interesting properties. Though qualitatively 
the nature of all equilibrium points are similar to that of dust, due to vanishing 
of the trace of the energy momentum tensor for standard matter, the surface 
defined by $x=0$ 
becomes an invariant subspace, containing the global sink/source $\mathcal{A}_\eps$, global 
saddles $\mathcal{B}_\eps$  
and the Einstein static point $\mathcal{ES}$ which is also a saddle. 
By definition, $x=0$ corresponds to 
$\dot{R}=0$, and therefore the points $\mathcal{A}_\eps$ are not accelerated
expansion/collapse solutions but they are instead static Minkowski points. 
Figure 5 shows the phase portrait of this subspace.  It is interesting to note that in 
this subspace the points $\mathcal{B}_\eps$ behaves like source/sink, however in the complete 
state space they are saddles. All the trajectories from $\mathcal{A}_-$ to $\mathcal{A}_+$, 
and from $\mathcal{B}_+$ to $\mathcal{B}_-$ admit bouncing universes.   
Also figure 4 shows a solution for radiation like matter linking 
$\mathcal{N}_+$ to $\mathcal{L}_-$ which has a bounce in the future.

\begin{table}[bp] \centering
\caption{Equilibrium points of the FRW state space. We have abbreviated $k(n)=\sfrac{n-2}{3(n-1)(2n-1)}$,
$j(n)=(n-3)(n-1)^2$ and $P_+=(1+\sqrt{3})/2\approx 1.37$. The quantities $a_0,a_1,a_2,a_3$ 
are constants of integration.}

\begin{tabular}{ll|l|l}
\multicolumn{3}{c}{}\\
\br Point  & $(Q,x,y)$ & constraints & Solution/Description
\\
& & & \\\br
& & & \\

$\mathcal{N}_\eps$& $\left(0,\;\eps,\;0\right)$ & $n\in [1,3]$& Vacuum Minkowski\\
& & &\\
\hline\\
$\mathcal{L}_\eps$  & $\left(2\eps,\;-\eps,\;0\right)$ & $n\in
[1,3]$ &
Vacuum Minkowski\\
& & &\\
\hline
$\mathcal{B}_\eps$ & $\left(\sfrac{\eps}{3-n}\,\;\eps\sfrac{n-2}{n-3},\;0\right)$
& $n\in[1,2.5]$ & Vacuum Minkowski\\
& & &\\
\hline
& & & Vacuum, Flat, Acceleration$\neq0$\\ 
$\mathcal{A}_\eps$ &
$\left(\eps\sfrac{2n-1}{3(n-1)},\eps\sfrac{n-2}{3(n-1)},\sfrac{8n^2-14n+5}{9(n-1)^2}\right)$ 
&$n\in[1.25,3]$  & Decelerating for $P_+<n<2$\\
& & & $a(t)=a_0\left(a_1+k(n)t\right)^{-3k(n)}$\\
& & &\\
\hline
Line  &
 & $|Q|\le\sfrac{1}{2-n}$ for $n\in [1,P_+]$ & 
Non-Accelerating curved\\
$\mathcal{LC}$ &$\left(Q,\;-Q(n-1),\;\sfrac{j(n)Q^2+n-1}{n}\right)$ &  $|Q|\le \sfrac{1}{\sqrt{3}(n-1)}$ for $n \in [P_+,3]$ & $a(t)=a_2t+a_3$\;\;, $\rho^m(t)>0$\\
 & & &\\
 \hline

\br
\end{tabular}\label{tab:eq-points}
\end{table}

\begin{table}[tbp] \centering
\caption{This table summarizes the nature of the \equi
points. As in the previous table, we have
abbreviated $P_+=(1+\sqrt{3})/2\approx 1.37$. 
The physically interesting values correspond to $n=3/2$ (dust) and $n=2$ (radiation). Note for $n=3/2$, the Einstein Static point becomes a bifurcation, and its stability cannot be determined within the linear theory (see text).} 
\begin{tabular}{c|c|ccccc|}
\multicolumn{6}{c}{} \\ \br
 \footnotesize{point}& type & \multicolumn{5}{c}{\footnotesize{range of n}}\\
 \cline{3-7}
 &&
  \footnotesize{$(1,5/4)$} & \footnotesize{$(5/4,P_+)$} &
\footnotesize{$(P_+,3/2)$} & \footnotesize{$(3/2,5/2)$}& \footnotesize{$(5/2,3)$}\\
\hline

\hline \hline

\footnotesize{$\mathcal{A}_+$}&\footnotesize{expanding}&\multicolumn{1}{c|}{\footnotesize{--}}
& \multicolumn{1}{c|}{\footnotesize{saddle}} &
\multicolumn{3}{c|}{\footnotesize{{\bf sink} }} \\
\hline

\footnotesize{$\mathcal{A}_-$}&\footnotesize{collapsing}&\multicolumn{1}{c|}{\footnotesize{--}}
& \multicolumn{1}{c|}{\footnotesize{saddle}} &
\multicolumn{3}{c|}{\footnotesize{{\bf source} }} \\
\hline

\hline \small{$\mathcal{B}_\pm$}&\footnotesize{static}&
\multicolumn{4}{c|}{\footnotesize{saddle}}&\multicolumn{1}{c|}{\footnotesize{--}} \\
\hline

 \small{$\mathcal{L}_+$} &\footnotesize{static} &
\multicolumn{1}{c|}{\footnotesize{saddle}}
& \multicolumn{3}{c|}{\footnotesize{{\bf source}}} &\multicolumn{1}{c|}{\footnotesize{saddle}} \\

\hline  \small{$\mathcal{L}_-$} &\footnotesize{static} &
\multicolumn{1}{c|}{\footnotesize{saddle}}
& \multicolumn{3}{c|}{\footnotesize{{\bf sink}}} &\multicolumn{1}{c|}{\footnotesize{saddle}} \\

\hline \small{$\mathcal{N}_+$}&\footnotesize{static}&
\multicolumn{5}{c|}{\footnotesize{{\bf source}}} \\

\hline \small{$\mathcal{N}_-$}&\footnotesize{static}&
\multicolumn{5}{c|}{\footnotesize{{\bf sink}}} \\

\hline
\hline \small{ $\mathcal{LC}^{\rm exp}$ } &\footnotesize{expanding} & \multicolumn{2}{c|}{\footnotesize{{\bf sink}}}&
\multicolumn{1}{c|}{\footnotesize{{\bf sink}  \, (for $Q<Q_b$)} }& \multicolumn{2}{c|}{\footnotesize{saddle}} \\
&&&&
\multicolumn{1}{|c|}{\footnotesize{saddle (for $Q>Q_b$)}} &  \\
\hline \small{ $\mathcal{LC}^{\rm coll}$} &\footnotesize{collapsing}& \multicolumn{2}{c|}{\footnotesize{{\bf source}}}&
\multicolumn{1}{c|}{\footnotesize{{\bf source} (for $|Q|<Q_b$)}} & \multicolumn{2}{c|}{\footnotesize{saddle}} \\

&&&&
\multicolumn{1}{|c|}{\footnotesize{saddle \,\,(for $|Q|>Q_b$)}} &  \\

\hline
\hline \hline
\end{tabular}\label{tab:nature-B1}
\end{table}

\section{Discussion and Conclusions}
In this work, we have studied the state space of the class of isotropic FRW models in $f(R)$-gravity. We were particularly interested in the stability of the Einstein Static model, which in general only exists for the specific form of $f(R)=R^n$, with $n$ constrained to be a function of the equation of state parameter $w$ as shown in \cite{Goswami:2008fs}. For this reason, we here considered $R^n$-gravity with the constraint $n=3/2(1+w)$ only. 

We have identified the past and future attractors of the state space of these models. We have found that the \ES model is an unstable saddle point for all equations of state stiffer than dust ($w>0$ or $n>3/2$), and a neutrally stable center for $w\in(-1/3,0)$, i.e. $n\in(1,3/2)$. This is different to General Relativity, where the \ES model is an unstable saddle for all $w\in(-1/3,1)$. We have also numerically found bouncing orbits that link the decelerating collapsing point $\mathcal{A}_-$ to the accelerating expanding point $\mathcal{A}_+$ via an asymptotically Einstein static phase represented by the saddle point $\mathcal{ES}$.

We note that the stability of the Einstein Static model obtained here is in agreement with the analysis of inhomogeneous perturbations \cite{Goswami:2008fs}, provided we carefully examine the relationship between the dynamical systems analysis here and the harmonically decomposed linear perturbations described in \cite{Goswami:2008fs}. The homogeneous mode with comoving wavenumber $\nt=0$ was excluded in \cite{Goswami:2008fs}, since it corresponds to a change in the background of the linear perturbation theory. It is this mode however that corresponds to the stability properties of the dynamical system examined here. This mode is neutrally stable if  $w<0$, and unstable with a decaying and a growing mode if $w>0$. These results exactly match the results from the dynamical systems analysis 
presented in this paper. \\
\par
\noindent{\bf Acknowledgments}\\
We thank Sante Carloni and Kishore Ananda for valuable input and the National Research Foundation (South Africa) for financial support.

\section*{References}

\end{document}